# New Method to Improve the Accelerating Gradient of Superconducting Cavity*

ZhenChao Liu[#], Jie Gao

Institute of High Energy Physics, Chinese Academy of Sciences, Beijing 100049, China

Abstract:

Quench is a common phenomenon in a superconducting cavity and often limits the accelerating gradient of the cavity. Accurate location of the quench site can be located by second sound detection. For multi-cell superconducting cavity, one defect may cause the cell with defect quenches and then the whole cavity quenches. Now we proposed a new method to eliminate the bad influence of the quench cell to the whole cavity.

## 1. Introduction

Quench is a common phenomenon in a superconducting cavity and often limits the accelerating gradient of the cavity. Quench also typically limits the maximum achievable surface magnetic field to a value less than the thermodynamic critical field (~200 mT [1, 2]). Accurate location of the quench site can be located by second sound detection [3-6]. Actually, the maximum accelerating gradient of the superconducting cavity is limited by the quench cell. The reason is that the field in each cell is tuned to the same. People used different method to eliminate quench in the cavity. And the cavity accelerating gradient can be improved after the quench cell field increased. In this paper, we proposed a new method to improve the accelerating gradient by just adjusting the field in each cell.

## 2. The New Method

For cavity with cell number $n$ and same cell length, the accelerating gradient $E_{acc}$ for the whole cavity is

$$E_{acc} = \frac{1}{n}\sum_{1}^{n} E_{acc,i} \qquad (1)$$

Here, $E_{acc,i}$ is the accelerating gradient for each cell.

Usually we tune the field flatness to >97% in the cavity. That makes the cavity accelerating gradient limited by the worse cell. However, other cell may reach much higher field. We can improve the accelerating gradient by making each cell reach its maximum field. This is the opposite direction of field flatness tuning. The result is that the cell field is not flat.

For the situation in figure 1, the left end cell is the quench cell and the accelerating gradient can only reach 20 MV/m, the accelerating gradient of other cells can reach 30MV/m. For the usual field flatness tuning (red line), this cavity can only reach 20MV/m. If we tune the cavity field flatness to the shape of the green line in figure 1, the whole cavity accelerating gradient can be improved to about 28MV/m.

*Project 11275226 and 11175192 supported by NSFC
#zcliu@ihep.ac.cn

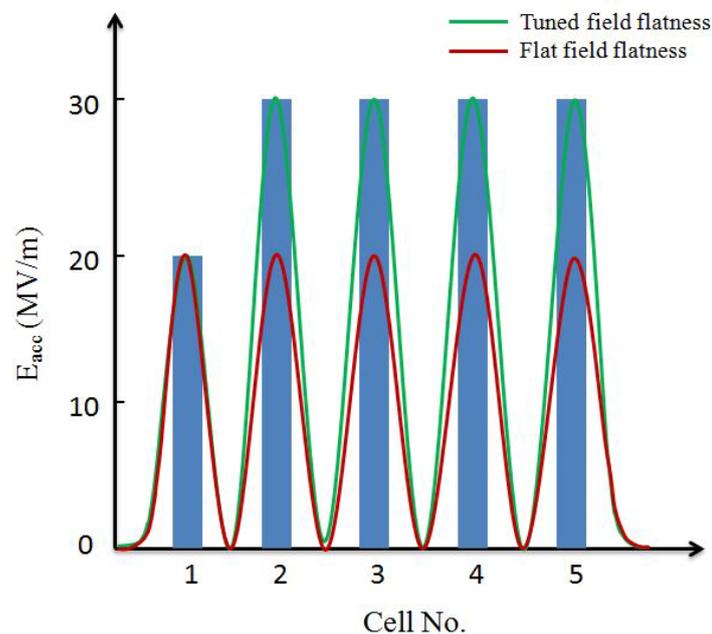

FIG. 1. Cavity maximum cell field and field tuning method.

Figure 1 shows the simplest situation. There is only one cell field lower than others and the other cells have the same maximum field. The real situation is more complicated. However, we can use this tuning method to all the other situations. That is tuning the field level lower in the worst cell and keeping the field level equal for other cells. The proportional of the tuning field level should be the same as the maximum field in the worse cell and the lowest field of other cells. Figure 2 shows the tuning method for other situations.

The ideal tuning is that all the fields in each cell are tuned to the maximum of the cell. However, this is very complicated. We can apply the upper method to improve the accelerating gradient of the cavity.

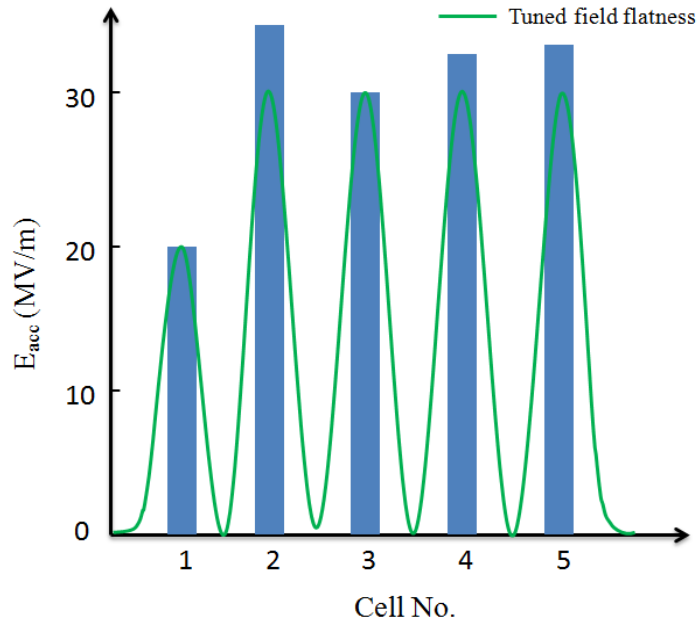

FIG. 2. Cavity maximum cell field and field tuning method.

This method can be easily applied to cavities with different cell numbers. The accelerating gradient of ILC cavity should be larger than 31.5MV/m. It is a challenging objective. The cavity has nine cells and each cell should be reaching 31.5MV/m. However, by using this method, the gradient of the cavity with one or two bad cell can be improved. From the practice experience of tuning the ILC cavity IHEP02 in IHEP, it is easy to tune the field shape with one lower and the others same.

## 3. Tuning Procedure

The cavity should be tuned flat first and measured in the vertical test. We can get the accelerating gradient and find the bad cell if quench happens. Then the maximum field in each cell needs to be measured. The maximum field of each cell of the cavity can be measured in the vertical test by passband mode test of TM010 [7, 8]. After the first vertical test, the cavity field flatness should be tuned as the bad cell lower and other cells the same. After that, we can do the second vertical test and measure the accelerating gradient improvement.

The field flatness will slightly change when the cavity was put in the cryomodule [9]. However, it will not decrease the performance of the tuned cavity too much. This method can significantly improve the accelerating gradient of the cavity with very low field cell.

## 4. Conclusion

The accelerating gradient is the key parameter of superconducting cavity. In most applications such as ILC, the accelerating gradient is much higher much better. However, the accelerating gradient is often limited by quench. People usually improve the gradient by post processing such as BCP, EP, HPR, etc.. We proposed a new method by just tuning the field flatness shape and it will significantly improve the accelerating gradient for the cavity with one or two bad cells.

## 4. References

1. H.R. Kerchner, D.K. Christen and S.T. Sekula, Critical field Hc and Hc2 of superconducting niobium, Physical Review B, Vol 24,No 3.
2. J. Daams and J.P. Carbotte, Thermodynamic properties of superconducting niobium, Journal of Low Temperature Physics, Vol. 40, Nos. 1/2,1980.
3. Z. A. Conway et al., OSCILLATING SUPERLEAK TRANSDUCERS FOR QUENCH DETECTION IN SUPERCONDUCTING ILC CAVITIES COOLED WITH HE-II, in Proceedings of Linac08, Victoria, BC, Canada, 2008, THP036.
4. M.P. Kelly, M. Kedzie and Z. Liu, A SIMPLE SECOND SOUND DETECTION TECHNIQUE FOR SRF CAVITIES, in Proceedings of SRF2009, Berlin, Germany, 2009, TUPPO032.
5. K. Liao et al., Second sound measurement for SPL cavity diagnostics, in Proceedings of SRF2011, Chicago, IL USA, 2011, THPO026.
6. G. Eremeev, COMMISSIONING CORNELL OSTS FOR SRF CAVITY QUENCH IDENTIFICATION AT JLAB, in Proceedings of SRF2011, Chicago, IL USA, 2011, THPO029.
7. J. Gao et al., FIRST TEST RESULT OF THE IHEP-01 LARGE GRAIN 9-CELL CAVITY, in Proceedings of Linac10, Tsukuba, Japan, 2010, THP120.
8. K. Umemori et al., Results of vertical tests for KEK-ERL 9-cell superconducting cavity, in Proceedings of IPAC10, Kyoto, Japan, 2010, WEPEC030.
9. S. An and H. Wang, Tuner Effect on the Field Flatness of SNS Superconducting RF Cavity, JLAB-TN-03-043 or SNS-NOTE-AP119 (2003).